\documentclass[journal]{IEEEtran}

\usepackage{booktabs}
\usepackage{threeparttable}
\usepackage{amsmath,graphicx,cite,amssymb}
\usepackage{amsfonts,multirow,bm,array,setspace,stfloats}

\usepackage{graphicx,float,cite,amssymb}
\usepackage{graphics}
 \DeclareGraphicsExtensions{.pdf,.jpeg,.png,.jpg}   
\usepackage{multirow,bm,bbm,array,setspace}
\usepackage{textcomp}
\usepackage{psfrag}
\usepackage{color}
\usepackage{pstricks,enumerate}
\usepackage{bbm}
\usepackage{subfigure}

\usepackage{algorithm,algpseudocode}
\makeatletter
\newcommand{\distas}[1]{\mathbin{\overset{#1}{\kern\z@\sim}}}%
\newsavebox{\mybox}\newsavebox{\mysim}
\newcommand{\distras}[1]{%
  \savebox{\mybox}{\hbox{\kern1pt$\scriptstyle#1$\kern1pt}}%
  \savebox{\mysim}{\hbox{$\sim$}}%
  \mathbin{\overset{#1}{\kern\z@\resizebox{\wd\mybox}{\ht\mysim}{$\sim$}}}%
}
\makeatother
\ifCLASSINFOpdf
\else
\fi

\newtheorem{proposition}{Proposition}

\newtheorem{theorem}{Theorem}

\newtheorem{example}{Example}

\newtheorem{remark}{Remark}

\newcommand{\bA}{\bm A}
\newcommand{\ba}{\bm a}

\newcommand{\bC}{\bm C}

\newcommand{\bx}{\bm{x}}

\newcommand{\bv}{\bm v}
\newcommand{\bV}{\bm V}

\newcommand{\by}{\bm{y}}

\newcommand{\bh}{\bm{h}}

\newcommand{\bw}{\bm{w}}

\newcommand{\btheta}{\bm \theta}

\begin{document}
%
\title{Configuring Intelligent Reflecting Surface with Performance Guarantees: Optimal Beamforming}
\author{
	\IEEEauthorblockN{
	Yaowen Zhang, \IEEEmembership{Student Member,~IEEE},
	Kaiming Shen, \IEEEmembership{Member,~IEEE},\\
	Shuyi Ren, \IEEEmembership{Student Member,~IEEE},
	Xin Li, Xin Chen, Zhi-Quan Luo, \IEEEmembership{Fellow,~IEEE}
} 
\thanks{Manuscript received \today. This work was supported in part by  the National Natural Science Foundation of China (NSFC) under Grant 62001411 and in part by the Huawei Technologies. This article was presented in part at the IEEE Global Communications Conference (GLOBECOM), Madrid, Spain, 2021.

Y. Zhang, K. Shen, and S. Ren are with the School of Science and Engineering, The Chinese University of Hong Kong, Shenzhen, 518172, China (e-mail: yaowenzhang@link.cuhk.edu.cn; shenkaiming@cuhk.edu.cn; shuyiren@link.cuhk.edu.cn). 

X. Li and X. Chen are with Huawei Technologies (e-mail: razor.lixin@huawei.com; chenxin@huawei.com).

Z.-Q. Luo is both with the School of Science and Engineering, The Chinese University of Hong Kong, Shenzhen, 518172, China and with Shenzhen Research Institute of Big Data, China (e-mail: luozq@cuhk.edu.cn).}
}

%


\maketitle

\begin{abstract}
	This work proposes linear time strategies to optimally configure the phase shifts for the reflective elements of an intelligent reflecting surface (IRS). Specifically, we show that the binary phase beamforming can be optimally solved in linear time to maximize the received signal-to-noise ratio (SNR). For the general $K$-ary phase beamforming, we develop a linear time approximation algorithm that guarantees performance within a constant fraction $(1+\cos(\pi/K))/2$ of the global optimum, e.g., it can attain over $85\%$ of the optimal performance for the quadrature beamforming with $K=4$. According to the numerical results, the proposed approximation algorithm for discrete IRS beamforming outperforms the existing algorithms significantly in boosting the received SNR.
\end{abstract}


\section{Introduction}
\label{sec:overview}

\IEEEPARstart{I}{ntelligent} reflecting surface (IRS) is an emerging 6G technology that uses a large array of low-cost electromagnetic ``mirrors'', i.e., reflective elements, to orient the impinging radio waves toward the intended receiver, thereby boosting the spectral efficiency, energy efficiency, and reliability of wireless transmission \cite{wu_zhang_MCOM,debbah_TWC19}. From a practical standpoint, the choice for the phase shift induced by each reflective element is normally restricted to a set of discrete values because of hardware constraints. Although IRS has been studied extensively in the literature to date, it remains a challenging task to optimally coordinate discrete phase shifts across reflective elements.

This work proposes efficient strategies for discrete beamforming for IRS in order to maximize the received signal-to-noise ratio (SNR) boost, the computational complexity of which is only linear in the number of reflective elements. For the binary phase beamforming with each phase shift confined to the set $\{0,\pi\}$, we show that the global optimal solution can be computed in linear time; for the general $K$-ary phase beamforming case with each phase shift confined to the set $\{\omega,2\omega,\ldots,K\omega\}$ where $\omega=2\pi/K$, we propose a linear time approximation algorithm that guarantees an SNR boost within a constant fraction $(1+\cos(\pi/K))/2$ of the global optimum, e.g., it can attain over $85\%$ of the optimal SNR boost for the quadrature beamforming with $K=4$.



Unlike the existing multi-antenna devices, IRS is a passive device that does not perform active signal transmitting. This passive trait of IRS has spurred considerable research interests over the past few years in adapting the traditional methods of active beamforming, covering a variety of system models ranging from point-to-point communication \cite{fang_SVD_BF_20,shuowen_double_BF_20,poor_active_BF_21,Ng_impairment_BF_21} to downlink cellular network \cite{rui_SDP_BF_19,Larsson_FP_BF_19,Ding_Pham_BF_20,Dhahir_SCA_BF_20,Ni_FP_BF_20,Nallanathan_SDP_BF_20,Pei_FP_BF_21,rui_double_BF_21,YF_maxmin_BF_21,mm_2timescale_BF_21,Ding_FP_BF_21,Dai_FP_BF_21,Skoglund_decentralized_BF_21}, uplink cellular network \cite{Dobre_SDP_BF_21,Ni_Lin_FP_BF_21}, interference channel \cite{Shi_FP_BF_21}, and wiretap channel \cite{Chen_FP_BF_21}. While the prior studies in this area mostly assume that the phase shift can be chosen arbitrarily for every reflective element, there is a growing trend toward practical beamforming that restricts the choice for phase shift to a set of discrete values \cite{Gao_DBF_21,Qian_DBF_21,you_zheng_zhang_jsac20,Schober_DBF_21,Smith_DBF_21,Yuen_DBF_20,wu_zhang_TCOM20,Balakrishnan_NSDI20,cai_li_COML20,di_song_jsac20}. 
Two common objectives of IRS beamforing as adopted in the literature are to maximize the throughput \cite{Larsson_FP_BF_19,fang_SVD_BF_20,Ni_FP_BF_20,Dhahir_SCA_BF_20,Dobre_SDP_BF_21,Dai_FP_BF_21,Skoglund_decentralized_BF_21,YF_maxmin_BF_21,Pei_FP_BF_21,Chen_FP_BF_21} and to minimize the power consumption under the quality-of-service (QoS) constraints \cite{rui_SDP_BF_19,Nallanathan_SDP_BF_20,Ni_Lin_FP_BF_21,Ding_FP_BF_21,mm_2timescale_BF_21}, other works accounting for the generalized degree-of-freedom (GDoF) \cite{Shi_FP_BF_21}, energy efficiency \cite{Ding_Pham_BF_20}, and outage probability \cite{Hossain_FP_BF_21}. 


Optimization is a key aspect of the IRS beamforming. In the realm of continuous IRS beamforming, two standard optimization tools---semidefinite relaxation (SDR) \cite{luo_SDR} and  fractional programming (FP) \cite{FP1,FP2}---have been brought to the fore to address the nonvexity of the IRS beamforming problem. For instance, \cite{rui_SDP_BF_19,Johansson_est_19,rui_double_BF_21,Nallanathan_SDP_BF_20,Dobre_SDP_BF_21,YF_maxmin_BF_21,Skoglund_decentralized_BF_21} suggest using SDR because of the quadratic programming form of the IRS beamforming problem, while \cite{Ni_FP_BF_20,Chen_FP_BF_21,Ding_FP_BF_21,Hossain_FP_BF_21,Ni_Lin_FP_BF_21,poor_active_BF_21,Dai_FP_BF_21,Chen_FP_BF_21,Pei_FP_BF_21} invoke FP to deal with the fractional function of the signal-to-interference-plus-noise ratio (SINR).
Moreover, \cite{Shi_FP_BF_21} develops a unified Riemannian conjugate gradient algorithm to enable interference alignment in an IRS-assisted system, \cite{Dhahir_SCA_BF_20,mm_2timescale_BF_21} cope with the nonconvex beamforming problem via successive convex approximation, \cite{Ng_impairment_BF_21} relies on the minorization-maximization (MM) to address the hardware impairments in the IRS beamforming problem, and \cite{fang_SVD_BF_20,Skoglund_decentralized_BF_21} suggest a decentralized beamforming scheme based on alternating direction method of multipliers (ADMM). 

In comparison to the above continuous beamforming algorithms for IRS, the existing discrete beamforming algorithms lag somewhat in sophistication. Few attempts have been made for the discrete case to date. To achieve the global optimum, \cite{cai_li_COML20} uses the exhaustive search and \cite{di_song_jsac20} uses the branch-and-bound algorithm, neither of which is computationally scalable. The remaining works \cite{Gao_DBF_21,Qian_DBF_21,you_zheng_zhang_jsac20,Schober_DBF_21,Smith_DBF_21,Yuen_DBF_20,wu_zhang_TCOM20} typically relax the discrete constraint and then round the relaxed continuous solution to the closest point in the discrete constraint set, namely the closest point projection algorithm. However, this heuristic method can lead to arbitrarily bad performance as shown in this paper. We shall give a linear time beamforming algorithm with provable performance guarantees. The main contributions of the paper are two-fold:

\subsubsection{Binary phase beamforming}
We consider maximizing the SNR boost when each phase shift is selected from $\{0,\pi\}$. Although the corresponding problem appears intractable, we show that the global optimum can be achieved in linear time. In addition, we show that the common heuristic of projecting the relaxed continuous solution to the closest point in the discrete constraint set \cite{Gao_DBF_21,Qian_DBF_21,you_zheng_zhang_jsac20,Schober_DBF_21,Smith_DBF_21,Yuen_DBF_20,wu_zhang_TCOM20,Balakrishnan_NSDI20} can result in arbitrarily bad performance.

\subsubsection{General $K$-ary phase beamforming}
We then assume that each phase shift is selected from a fixed set of $K$ discrete values. The proposed linear time algorithm has provable performance of reaching the global optimum to within a constant fraction $1/2+\cos(\pi/K)/2$, thus guaranteeing a tighter approximation ratio than the closest point projection algorithm \cite{Gao_DBF_21,Qian_DBF_21,you_zheng_zhang_jsac20,Schober_DBF_21,Smith_DBF_21,Yuen_DBF_20,wu_zhang_TCOM20,Balakrishnan_NSDI20} and the SDR algorithm \cite{rui_SDP_BF_19,Johansson_est_19,rui_double_BF_21,Nallanathan_SDP_BF_20,Dobre_SDP_BF_21,YF_maxmin_BF_21,Skoglund_decentralized_BF_21}.

The above results are of theoretical significance because they shatter the common beliefs that the discrete IRS beamforming problem must be NP-hard and that the closest point projection is the best strategy for coordinating phase shifts in practice. Furthermore, it is worth mentioning that the proposed algorithm in this work can be extended to \emph{blind beamforming} without channel estimation. This topic is pursued in the companion paper \cite{IRS:blind_21} to the present.




Throughout the paper, we use the boldface lower-case letter to denote a vector, the bold upper-case letter a matrix, and the calligraphy upper-case letter a set. For a matrix $\bA$, $\bA^\mathsf{T}$ refers to the transpose, $\bA^\mathsf{H}$ the conjugate transpose, and $\bA^{-1}$ the inverse. For a vector $\ba$, $\|\ba\|$ refers to the Euclidean norm, $\ba^\mathsf{T}$ the transpose, and $\ba^\mathsf{H}$ the conjugate transpose. The cardinality of a set $\mathcal Q$ is denoted as $|\mathcal Q|$. The set of real numbers is denoted as $\mathbb R$. The set of complex numbers is denoted as $\mathbb C$. For a complex number $u\in\mathbb C$, $\mathrm{Re}\{u\}$, $\mathrm{Im}\{u\}$, and $\mathrm{Arg}(u)$ refer to the real part, the imaginary part, and the principal argument of $u$, respectively. 
The rest of the paper is organized as follows. Section \ref{sec:model} describes the system model and problem formulation. Section \ref{sec:binary} discusses binary phase beamforming. Section \ref{sec:K-ary} discusses $K$-ary phase beamforming. Section \ref{sec:simulations} presents the numerical results. Section \ref{sec:conclusion} concludes the paper.

\section{System Model}
\label{sec:model}

Consider a pair of transmitter and receiver, along with an IRS that facilitates the data transmission between them. The IRS consists of $N$ passive reflective elements. Let $h_n\in\mathbb C$, $n=1,\ldots,N$, be the cascaded channel from the transmitter to the receiver that is induced by the $n$th reflective element; let $h_0\in\mathbb C$ be the superposition of all those channels from the transmitter to the receiver that are not related to the IRS, namely the \emph{background channel}.  Each channel can be rewritten in an exponential form as
\begin{equation}
\label{hn}
	h_n= \beta_ne^{j\alpha_n},\;n=0,\ldots,N,
\end{equation} 
with the magnitude $\beta_n\in(0,1)$ and the phase $\alpha_n\in[0,2\pi)$.

Denote the IRS beamformer as $\bm\theta=(\theta_1,\ldots,\theta_N)$, where each $\theta_n\in[0,2\pi)$ refers to the phase shift of the $n$th reflective element. The choice of each $\theta_n$ is restricted to the discrete set
\begin{equation}
\label{Phi}
	\Phi_K= \big\{\omega,2\omega,\ldots,K\omega\big\}
\end{equation}
with the distance parameter
\begin{equation}
\label{omega}
	\omega=\frac{2\pi}K.
\end{equation}
Let $X\in\mathbb C$ be the transmit signal with the mean power $P$, i.e., $\mathbb E[|X|^2]=P$. The received signal $Y\in\mathbb C$ is given by
\begin{align}
    Y &= \Bigg(h_0+\sum^N_{n=1}h_ne^{j\theta_n}\Bigg)X+Z,
\end{align}
where an i.i.d. random variable $Z\sim\mathcal{CN}(0,\sigma^2)$ models the additive thermal noise. The received SNR can be computed as
\begin{subequations}
\begin{align}
\label{snr}
	\mathsf{SNR} &=\frac{\mathbb E[|Y-Z|^2]}{\mathbb E[|Z|^2]}\\
	&=\frac{P\Big|\beta_0e^{j\alpha_0}+\sum^N_{n=1}\beta_ne^{j(\alpha_n+\theta_n)}\Big|^2}{\sigma^2}.
\end{align}
\end{subequations}
The baseline SNR without IRS is
\begin{equation}
	\mathsf{SNR}_0 = \frac{P\beta^2_0}{\sigma^2}.
\end{equation}
The \emph{SNR boost} is defined as
\begin{subequations}
\begin{align}
\label{SNR_boost}
	f(\bm\theta) &= \frac{\mathsf{SNR}}{\mathsf{SNR}_0}\\
	&=\frac{1}{\beta^2_0}\left|\beta_0e^{j\alpha_0}+\sum^N_{n=1}\beta_ne^{j(\alpha_n+\theta_n)}\right|^2.
\end{align}
\end{subequations}
We seek the optimal $\bm\theta$ to maximize the SNR boost:
\begin{subequations}
\label{prob:snr}
\begin{align}
\underset{\bm\theta}{\text{maximize}} &\quad f(\bm\theta)
    \label{prob:snr:obj}\\
\text{subject to}&\quad \theta_n\in\Phi_K,\;\forall n=1,\ldots,N.
    \label{prob:snr:cons}
\end{align}    
\end{subequations}
The above problem is difficult to tackle directly because of the discrete constraint on $\bm\theta$.



\section{Binary Phase Beamforming with $K=2$}
\label{sec:binary}



As $K\rightarrow\infty$, the discrete beamforming problem in (\ref{prob:snr}) reduces to the continuous, in which case the SNR boost is maximized when every $h_n$ is perfectly aligned with $h_0$, so the optimal relaxed solution $\theta^\infty_n$ equals the phase difference between $h_0$ and $h_n$:
\begin{equation}
\label{continuous_solu}
	\theta^\infty_n = \alpha_0-\alpha_n.
\end{equation}
It is then tempting to believe that the problem becomes harder when $K$ decreases. But it turns out that the problem can be efficiently solved when $K=2$, as stated below.
\begin{theorem}
\label{prop:optimal_binary_BF}
	The binary phase beamforming problem in (\ref{prob:snr}) with $K=2$ can be optimally solved in $O(N)$ time.
\end{theorem}

Before proceeding to the proof of the above theorem, we first cast (\ref{prob:snr}) to a quadratic programming form. By letting
\begin{equation}
\label{x:vec}
	\bx= (1,x_1,\ldots,x_N)^{\mathsf{T}}\;\;\text{with}\;\;x_n = e^{j\theta_n},
\end{equation}
we rewrite the problem in (\ref{prob:snr}) as
\begin{subequations}
\label{prob:quadratic:complex}
\begin{align}
\underset{\bx}{\text{maximize}} &\quad \bx^{\mathsf{H}}\bC\bx
    \label{prob:quadratic:complex:obj}\\
\text{subject to}&\quad x_n\in\{e^{j\omega},\ldots,e^{jK\omega}\},\;\forall n,
    \label{prob:quadratic:complex:constraint}
\end{align}    
\end{subequations}
where the matrix $\bC=[C_{mn}]_{(N+1)\times(N+1)}$ is given by
\begin{equation}
\label{C}
    C_{mn}=h^{\mathsf{H}}_mh_n,\;\forall m,n=0,\ldots,N.
\end{equation}
It can be shown that $\mathrm{rank}(\bC)\le2$. We deal with the rank-one case and the rank-two case separately in the following.

\subsubsection{$\mathrm{rank}(\bC)=1$}
In this case, each $h_n$ is a multiple of $h_0$, i.e., the quotient
\begin{equation}
\label{hn_multiple}
	r_n = \frac{h_n}{h_0}\in\mathbb R.
\end{equation}
When $K=2$, by substituting (\ref{hn_multiple}) back into (\ref{prob:quadratic:complex}), we obtain
\begin{subequations}
\label{prob:quadratic:real}
\begin{align}
\underset{\bx}{\text{maximize}} &\quad (1+r_1x_1+\ldots+r_Nx_N)^2
    \label{prob:quadratic:real:obj}\\
\text{subject to}&\quad x_n\in\{-1,1\},\;\forall n.
    \label{prob:quadratic:real:constraint}
\end{align}    
\end{subequations}
It can be readily seen that the optimal $\bx$ would let every term $r_nx_n$ be positive, i.e.,
\begin{equation}
\label{opt_x:rank1}
    x^\star_n=\mathrm{sgn}(r_n),\;n=1,\ldots,N,
\end{equation}
where $\mathrm{sgn}(z)$ equals $1$ if $z\ge0$ and equals $-1$ otherwise. We then recover the optimal phase shift as $\theta_n^\star=\arccos(x_n^\star)$.
The above method runs in $O(N)$ time.

\subsubsection{$\mathrm{rank}(\bC)=2$}
Rewrite each $h_n$ as a vector $\bv_n\in\mathbb R^2$:
\begin{equation}
\label{bv_n}
	\bv_n = 
	\left[\begin{matrix}
		\mathrm{Re}\{h_n\}\\
		\mathrm{Im}\{h_n\}
	\end{matrix}\right],\;\;n=0,\ldots,N,
\end{equation}
and write
\begin{equation}
	\bV = \big[\bv_0,\ldots,\bv_N\big].
\end{equation}
With an auxiliary variable $\by\in\mathbb R^2$,
we transform the objective function in (\ref{prob:quadratic:complex}) as follows:
\begin{subequations}
\begin{align}
\bx^{\mathsf{H}}\bC\bx
&= \|\bV\bx\|^2\\
&= \underset{\|\by\|=1}\max\;\|\bV\bx\|^2\cdot\|\by\|^2\\
&= \underset{\|\by\|=1}\max\;\big((\bV\bx)^{\mathsf{T}}\by\big)^2
    \label{Cauchy_Schwarz}\\
&= \underset{\|\by\|=1}\max\;\big(\bv_0^{\mathsf{T}}\by+x_1\bv^{\mathsf{T}}_1\by+\ldots+x_N\bv^{\mathsf{T}}_N\by\big)^2,
    \label{xvy}
\end{align}
\end{subequations}
where (\ref{Cauchy_Schwarz}) follows by the Cauchy-Schwarz inequality. Inspection of the new objective function in (\ref{xvy}) shows that the optimal choice of $x_n\in\{-1,1\}$ given $\by$ is to let $x_n\bv^\mathsf{T}_n\by$ have the same sign as $\bv^\mathsf{T}_0\by$, i.e.,
\begin{equation}
\label{opt_xy}
    x^\star_n=\mathrm{sgn}(\bv^{\mathsf{T}}_0\by\bv^{\mathsf{T}}_n\by),\;n=1,\ldots,N,
\end{equation}
so the joint optimization of $(\bx,\by)$ hinges on how to optimize $\by$.

\begin{figure}
\begin{minipage}[b]{1.0\linewidth}
	  \centering
	  \centerline{\includegraphics[width=6cm]{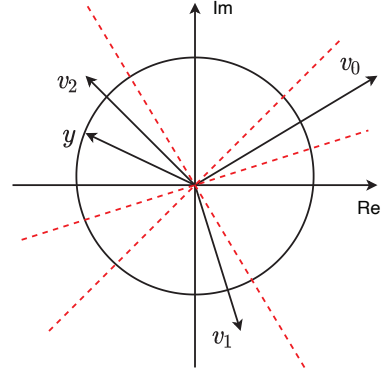}}
	  \vspace{0em}
\end{minipage}
\caption{The rank 2 case of binary phase beamforming with $N=2$. The normal vectors $\bv_0,\bv_1,\bv_2$ correspond to a tangent line each. The tangent lines, denoted as the dashed lines, partition the unit circle into 6 circular segments.}
\label{fig:binary_solu}
\end{figure}

We treat each $\bv_n$, $n=0,1,\ldots,N$, as a normal vector and consider its tangent line as shown in Fig.~\ref{fig:binary_solu}. Since multiple $\bv_n$'s may be associated with a common tangent line, the total number of tangent lines $M\le N+1$. Denote an arbitrary tangent line as $\ell_1$. Starting from $\ell_1$, denote the remaining $M-1$ tangent lines as $\ell_2,\ldots,\ell_M$ in counterclockwise order. Let $\mathcal V_m$ be the set of $\bv_n$'s associated with $\ell_m$, $m=1,\ldots,M$. As illustrated in Fig.~\ref{fig:binary_solu}, these tangent lines partition
the unit circle into a total of $2M$ circular segments. Choose a circular segment that first intercepts $\ell_1$ then intercepts $\ell_2$ counterclockwise; denote it as $\mathcal L_1$. The remaining circular segments are denoted as $\mathcal L_2,\ldots,\mathcal L_M$ in counterclockwise order. 

Notice that on each circular segment $x^\star_n$ in (\ref{opt_xy}) is fixed, so $\big(\bv_0^{\mathsf{T}}\by+x^\star_1\bv^{\mathsf{T}}_1\by+\ldots+x^\star_N\bv^{\mathsf{T}}_N\by\big)$ is a piecewise linear function of $\by$. When $\by$ is restricted to a particular segment $\mathcal L_m$, the optimization problem of $\by$ in (\ref{xvy}) can be rewritten as
\begin{align}
\label{prob:wy}
\underset{\by\in\mathbb R^2,\,\|\by\|=1}{\text{maximize}} &\quad \bw^\mathsf{T}_m\by
\end{align}    
with the coefficient vector
\begin{equation}
\label{w_m}
	\bw_{m}=\sum^N_{n=0}\mathrm{sgn}(\bv^{\mathsf{T}}_n\by')\bv_n,
\end{equation}
where $\by'\in\mathbb R^2$ is an arbitrary vector located in the interior of $\mathcal L_m$. A naive idea is to compute every $\bw_m$ by using (\ref{w_m}), requiring $O(N)$ time; this would result in $O(N^2)$ time in total for all $m=1,\ldots,M$. But we show that the time can be reduced to $O(N)$ via iterative updating. When updating $x^\star_{n}$ from one circular segment $\mathcal L_m$ to the next circular segment $\mathcal L_{m+1}$, we just need to change the values of those $x^\star_{n}$ variables related to the tangent line $\ell_{m+1}$ that separates the two circular segments, i.e., $\{x^\star_n:\forall n\in\mathcal V_{m+1}\}$, so $\bw_{m+1}$ can be computed iteratively based on $\bw_m$ as
\begin{equation}
\label{w_m:iterative}
	\bw_{m+1}=\bw_{m} - \sum_{n\in\mathcal V_{m+1}}2\cdot\mathrm{sgn}(\bv^{\mathsf{T}}_n\by')\bv_n,
\end{equation}
with an arbitrary $\by'\in\mathbb R^2$ located in the interior of $\mathcal L_m$. Thus, while $\bw_1$ is still computed as in (\ref{w_m}) in $O(N)$ time, the other coefficient vectors $\bw_2,\ldots,\bw_M$ can be obtained sequentially according to (\ref{w_m:iterative}) in $O\big(\sum^M_{m=2}|\mathcal V_{m}|\big)=O(N)$ time in total.

Moreover, the optimal $\by$ in problem (\ref{prob:wy}) for each $m$ can be obtained in $O(1)$ time as
\begin{equation}
\label{opt_ym}
	\by^\star_m = \arg\min_{\by\in\mathbb R^2,\,\|\by\|=1} \left\|\by-\frac{\bw_m}{\|\bw_m\|}\right\|,\;m=1,\ldots,M,
\end{equation}
i.e., $\by^\star_m=\bw_m/\|\bw_m\|$ if $\bw_m/\|\bw_m\|\in\mathcal L_m$; otherwise $\by^\star_m$ is located at one of the endpoints of the circular segment $\mathcal L_m$. As a result, it requires $O(N)$ time overall to solve a sequence of problems in (\ref{prob:wy}) for $m=1,\ldots,M$.

Finally, the actual optimal $\by^\star$ is given by
\begin{equation}
\label{opt_y}
	\by^\star=\max\{\by^\star_1,\ldots,\by^\star_M\}.
\end{equation}
We then compute the corresponding $x^\star_n$ as in (\ref{opt_xy}) and recover the optimal phase shift by $\theta^\star_n=\arccos(x^\star_n)$. The entire procedure requires $O(N)$ time in total. The proof of Theorem \ref{prop:optimal_binary_BF}
is then completed. We summarize the above steps in Algorithm \ref{alg:binary}.

We remark that the binary phase beamforming case is fairly special in that the variable $x_n$ in (\ref{x:vec}), either $-1$ or $1$, is real-valued. When $K>2$, $x_n$ is complex-valued in general and thus the above linear time algorithm no longer applies.

In the existing literature \cite{Gao_DBF_21,Qian_DBF_21,you_zheng_zhang_jsac20,Schober_DBF_21,Smith_DBF_21,Yuen_DBF_20,wu_zhang_TCOM20}, a common way of discrete beamforming is to round the relaxed continuous solution $\theta^\infty_n$ in (\ref{continuous_solu}) to the closest point in the discrete set $\Phi_K$, i.e.,
\begin{equation}
\label{greedy_alg}
	{\theta}^{\text{CPP}}_n = \arg\min_{\theta_n\in\Phi_K}\big|\theta_n-\theta^\infty_n\big|,
\end{equation}
referred to as the \emph{Closest Point Projection (CPP)} algorithm. The following example however shows that the above algorithm can result in arbitrarily bad performance when $K=2$.
\begin{example}
	Consider the binary phase beamforming with $K=2$. Assume that all the reflected channels have equal magnitudes, and that half of the reflected channels have the phase $\alpha_n=\alpha_0-\epsilon+\pi/2$ while the other half have the phase $\alpha_n=\alpha_0+\epsilon-\pi/2$. As $\epsilon\rightarrow0$, the closest point projection algorithm leads to $f(\btheta^{\text{CPP}})\rightarrow1$, i.e., IRS does not bring any improvements.
\end{example}

\begin{algorithm}[t]
\caption{Proposed Binary Phase Beamforming Method}
\label{alg:binary}
\begin{algorithmic}[1]
\State\textbf{input:} $\bh_0,\bh_1,\ldots,\bh_N$
\For{$n=0,1,\ldots,N$}
	\State let $\bv_n=\big[\mathrm{Re}\{h_n\},\mathrm{Re}\{h_n\}\big]^\mathsf{T}$
	\State treat $\bv_n$ as the normal vector and draw its tangent line
\EndFor
\State obtain $M\le N$ distinct tangent lines $\ell_1,\ldots,\ell_M$; let $\mathcal V_m$ be the set of $\bv_n$'s associated with $\ell_m$, $m=0,1,\ldots,M$
\State the unit circle is partitioned into $M$ circular segments $\mathcal L_1,\ldots,\mathcal L_M$ as shown in Fig.~\ref{fig:binary_solu}
\State compute $\bw_1$ as in (\ref{w_m})
\State compute $\by^\star_1$ as in (\ref{opt_ym})
\For{$m=2,\ldots,M$}
	\State compute $\bw_m$ as in (\ref{w_m:iterative})
	\State compute $\by^\star_m$ as in (\ref{opt_ym})
\EndFor
\State $\by^\star=\max\{\by^\star_1,\ldots,\by^\star_M\}$
\For{$n=1,\ldots,N$}
	\State $x^\star_n=\mathrm{sgn}(\bv^\mathsf{T}_0\by^\star\bv^\mathsf{T}_n\by^\star)$
	\State $\theta^\star_n=\arccos(x^\star_n)$
\EndFor
\State\textbf{output:} $\btheta^\star=(\theta^\star_1,\ldots,\theta^\star_N)$
\end{algorithmic}
\end{algorithm}

\begin{figure}
\begin{minipage}[b]{1.0\linewidth}
	  \centering
	  \centerline{\includegraphics[width=4.4cm]{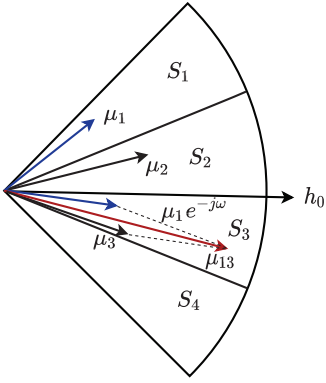}}
	  \vspace{0em}
\end{minipage}
\caption{The four sectors $\mathcal S_1$ to $\mathcal S_4$. In the proof of Theorem \ref{theorem:approx_ratio}, we first rotate $\mu_1$ clockwise by an angle of $\omega/2$, then combine it with $\mu_3$ to obtain $\mu_{13}$.}
\label{fig:proposed_alg}
\end{figure}

\section{General $K$-Ary Phase Beamforming}
\label{sec:K-ary}

This section considers the general $K$-ary phase beamforming in (\ref{prob:snr}). We begin with a sectorization scheme in order to reduce the search space of the optimal beamformer. This new idea then yields a linear time approximation algorithm that guarantees performance within a constant fraction of the global optimum. It is further shown that the proposed algorithm has a tighter approximation ratio than the existing algorithms.

\subsection{Optimal $K$-Ary Phase Beamforming}
\label{sec:approx_alg}

The following sectorization scheme is the building block of our approach to the $K$-ary phase beamforming. Consider four sectors around $h_0$ on the complex plane as illustrated in Fig.~\ref{fig:proposed_alg}:
\begin{multline}
\label{sectors}
	\mathcal S_i = \bigg\{u\in\mathbb C:\alpha_0+\frac{(2-i)\omega}{2}\le\mathrm{Arg}(u)\le\alpha_0+\frac{(3-i)\omega}{2}\bigg\},\\
	\forall i=1,2,3,4.
\end{multline}
The overall channel superposition under the optimal beamformer $\btheta^\star$ is denoted as
\begin{equation}
\label{h_sum}
g^\star = h_0+\sum^N_{n=1}h_ne^{j\theta_n^\star}.
\end{equation}
We remark that $\btheta^\star$ must satisfy
\begin{equation}
\label{arg:opt}
	\left|\mathrm{Arg}(h_0)-\mathrm{Arg}\left(\sum^N_{n=1}h_ne^{j\theta^\star_n}\right)\right|\le\frac{\omega}{2};
\end{equation} 
otherwise we could find an integer $k$ such that 
\begin{equation}
	\left|\mathrm{Arg}(h_0)-\mathrm{Arg}\left(\sum^N_{n=1}h_ne^{j(\theta^\star_n+k\omega)}\right)\right|\le\frac{\omega}{2},
\end{equation} 
and thus further increase the SNR boost. The bound in (\ref{arg:opt}) implies that $g^\star\in(\mathcal S_2\cup\mathcal S_3)$, so the possible values of $\btheta^\star$ are limited, as stated in the following proposition.


\begin{proposition}
\label{prop:opt_sectors}
	For the $K$-ary phase beamforming problem in (\ref{prob:snr}), the optimal solution $\btheta^\star$ is contained in either $\mathcal G(\mathcal S_1\cup\mathcal S_2\cup\mathcal S_3)$ or $\mathcal G(\mathcal S_2\cup\mathcal S_3\cup\mathcal S_4)$, where the function $\mathcal G(\cdot)$ is given by
	\begin{equation}
		\label{G}
		\mathcal G(\mathcal X) = \left\{\bm\theta\in\Phi^N_K\,\big|\,h_ne^{j\theta_n}\in\mathcal X, \;\forall n=1,\ldots,N\right\}
	\end{equation}
	with the subset $\mathcal X\subseteq\mathbb C$.
\end{proposition}
\begin{IEEEproof}
	If $g^\star$ is already known, then it is optimal to rotate every $h_n$ to the closest possible position to $g^\star$. As a result, every $h_ne^{j\theta^\star_n}$ must be in $\mathcal S_1\cup\mathcal S_2\cup\mathcal S_3$ if $g^\star\in\mathcal S_2$, and in $\mathcal S_2\cup\mathcal S_3\cup\mathcal S_4$ if $g^\star\in\mathcal S_3$. 
\end{IEEEproof}

In light of the above proposition, it suffices to search for $\bm\theta^\star$ in $\mathcal G(\mathcal S_1\cup\mathcal S_2\cup\mathcal S_3)\cup\mathcal G(\mathcal S_2\cup\mathcal S_3\cup\mathcal S_4)$, i.e.,
\begin{equation}
	\label{exact_alg}
	{\bm\theta}^\star = \arg\max_{\bm\theta\in\mathcal D}\, f(\bm\theta),
\end{equation}
where
\begin{equation}
	\label{D}
	\mathcal D= \mathcal G(\mathcal S_1\cup\mathcal S_2\cup\mathcal S_3)\cup \mathcal G(\mathcal S_2\cup\mathcal S_3\cup\mathcal S_4).
\end{equation}
Note that the search space size $|\mathcal D|=2^N$ is much smaller than the full solution space size $|\Phi^N_K|=K^N$ for large $K$. But the running time is still exponential. We aim at a more efficient algorithm in the next subsection.


\subsection{Proposed Approximation Algorithm}
\label{subsec:approximation_algorithm}

\begin{algorithm}[t]
\caption{Proposed $K$-Ary Phase Beamforming Method}
\label{alg:K-ary}
\begin{algorithmic}[1]
\State\textbf{input:} $\bh_0,\bh_1,\ldots,\bh_N$
\State $\mathcal D'=\emptyset$
\State compute the sectors $\mathcal S_1,\mathcal S_2,\mathcal S_3,\mathcal S_4$ as in (\ref{sectors})
\For{$i=1,2,3$}
	\State find $\bm\theta$ that lets every $h_ne^{j\theta_n}\in\mathcal S_i\cup\mathcal S_{i+1}$
	\State $\mathcal D'\rightarrow\mathcal D'\cup\{\bm\theta\}$
\EndFor
\State ${\btheta^{\text{APX}}} = \arg\max_{\bm\theta\in{\mathcal D'}}\, f(\bm\theta)$
\State\textbf{output:} $\btheta^{\text{APX}}$
\end{algorithmic}
\end{algorithm}

The premise behind the algorithm in (\ref{exact_alg}) and (\ref{D}) is that the optimally phase-shifted channels $\{h_1e^{j\theta_1^\star},\ldots,h_Ne^{j\theta_N^\star}\}$ can spread up to three consecutive sectors, i.e., $\mathcal S_1\cup\mathcal S_2\cup\mathcal S_3$ or $\mathcal S_2\cup\mathcal S_3\cup\mathcal S_4$. It guarantees the global optimality but results in the search space $\mathcal D$ in (\ref{D}) being exponentially large. 

To achieve a trade-off between complexity and performance, we propose letting the phase-shifted channels $\{h_1e^{j\theta_1},\ldots,h_Ne^{j\theta_N}\}$ spread at most two consecutive sectors, i.e., $\mathcal S_1\cup\mathcal S_2$ or $\mathcal S_2\cup\mathcal S_3$ or $\mathcal S_3\cup\mathcal S_4$. We now decide the IRS beamformer as
\begin{equation}
	\label{approx:theta}
	{\btheta^{\text{APX}}} = \arg\max_{\bm\theta\in{\mathcal D'}}\, f(\bm\theta)
\end{equation}
with a new search space
\begin{equation}
	\label{approx:D}
	{\mathcal D'} = \mathcal G(\mathcal S_1\cup\mathcal S_2)\cup \mathcal G(\mathcal S_2\cup\mathcal S_3)\cup \mathcal G(\mathcal S_3\cup\mathcal S_4).
\end{equation}
Algorithm \ref{alg:K-ary} summarizes the details.

Notice that $|\mathcal D'|\le3$, so the search in (\ref{approx:theta}) requires $O(1)$ time. Moreover, the new search space $\mathcal D'$ can be obtained in $O(N)$ time. Thus, the total running time is $O(N)$. Most importantly, the resulting SNR boost $f(\btheta^\text{APX})$ is within a constant fraction of the highest possible boost.

\begin{theorem}
	\label{theorem:approx_ratio}
	For the $K$-ary phase beamforming problem in (\ref{prob:snr}), the IRS beamformer $\btheta^\text{APX}$ in (\ref{approx:theta}) by the approximation algorithm  and the optimal IRS beamformer $\btheta^\star$ satisfy
	\begin{equation}
		\label{approx_ratio}
		  \frac{1+\cos(\pi/K)}{2} f(\btheta^\star) \le f(\btheta^{\text{APX}}) \le f(\btheta^\star).
	\end{equation}
\end{theorem}
\begin{IEEEproof}
Without loss of generality, assume that the optimal channel superposition $g^\star$ in (\ref{h_sum}) is located in $\mathcal S_2$. We partition $g^\star$ into three components:
\begin{equation}
	g^\star = \mu_1+\mu_2+\mu_3,
\end{equation}
where $\mu_i\in\mathbb C$ represents the superposition of those channels in $\{h_0,h_1e^{j\theta^\star_1},\ldots,h_Ne^{j\theta^\star_N}\}$ that are
located in $\mathcal S_i$. Defining an auxiliary variable (see Fig.~\ref{fig:proposed_alg} for illustration)
\begin{equation}
	\label{mu_13}
	\mu_{13}=\mu_1e^{-j\omega}+\mu_3,
\end{equation}
we can bound the value of $f(\bm\theta^\star)$ from above as
\begin{align}
	\label{bound:fstar}
	f(\btheta^\star) &= \frac{1}{\beta^2_0}\big|\mu_1+\mu_2+\mu_3\big|^2\nonumber\\
	&\le \frac{1}{\beta^2_0}\big(\big|\mu_{1}+\mu_{3}\big|+\big|\mu_2 \big|\big)^2\nonumber\\
	&\le \frac{1}{\beta^2_0}\big(\big|\mu_{13}\big| + \big|\mu_2\big|\big)^2,
\end{align}
where the last inequality follows since the rotation $e^{-j\omega}$ of $\mu_1$ in (\ref{mu_13}) reduces the angle between $\mu_1$ and $\mu_3$.

Because $\btheta^{\text{APX}}$ is located in either $\mathcal G(\mathcal S_1\cup \mathcal S_2)$ or $\mathcal G(\mathcal S_2\cup \mathcal S_3)$ when $g^\star\in\mathcal S_2$, we can lower bound the value of $f(\btheta^{\text{APX}})$ as
\begin{align}
	\label{bound:fhat}
	&f(\btheta^{\text{APX}})\notag\\
	&=\frac{1}{\beta^2_0}\cdot\max\big\{\big|\mu_1e^{-j\omega}+\mu_2+\mu_3\big|^2,\big|\mu_1+\mu_2+\mu_3e^{j\omega}\big|^2\big\}\nonumber\\
	&=\frac{1}{\beta^2_0}\cdot\max\big\{\big|\mu_{13}+\mu_2\big|^2, \big|\mu_{13}e^{j\omega}+\mu_2\big|^2\big\}\nonumber\\
	&\ge\frac{1}{\beta^2_0}\cdot\min_{|\mu'_2|=|\mu_2|}\max\big\{\big|\mu_{13}+\mu'_2\big|^2, \big|\mu_{13}e^{j\omega}+\mu'_2\big|^2\big\}\nonumber\\
	&= \frac{1}{\beta^2_0}\cdot\left(|\mu_2|^2+|\mu_{13}|^2+2\big|\mu_2\mu_{13}\big|\cos(\omega/2)\right),
\end{align}
where the last equality follows by fact that $\mu'_2=(|\mu_{2}|/|\mu_{13}|)\mu_{13}e^{j\omega/2}$ is the solution to the above min-max problem. With
$\lambda= |\mu_{13}|/|\mu_{2}|$,
we establish the following upper bound by combining (\ref{bound:fstar}) and (\ref{bound:fhat}):
\begin{align}
	\label{approx_ratio:proof}
	\frac{f({\btheta}^\star)}{f(\btheta^{\text{APX}})}
	&\le \frac{(\lambda+1)^2}{\lambda^2+1+2\lambda\cos(\omega/2)}
	\overset{(*)}{\le} \frac{2}{1+\cos(\omega/2)},
\end{align}
with equality in $(*)$ if and only if $\lambda=1$. Finally, plugging $\omega=2\pi/K$ in (\ref{approx_ratio:proof}) completes the proof.
\end{IEEEproof}

\begin{remark}
    The proposed algorithm can be extended to blind beamforming without any channel information as discussed in \cite{IRS:blind_21}. The work in \cite{IRS:blind_21} further proves that a polynomial number of random samples suffice to guarantee the optimal quadratic SNR boost.
\end{remark}

\subsection{Other $K$-Ary Phase Beamforming Algorithms}
\label{subsec:other_approximation_algorithm}

The proposed approximation algorithm is now compared with the existing methods for the $K$-ary phase beamforming. We begin with the closest point projection algorithm.

\begin{proposition}
\label{prop:greedy}
For the $K$-ary phase beamforming problem in (\ref{prob:snr}), the IRS beamformer $\btheta^{\text{CPP}}$ in (\ref{greedy_alg}) by the closest point projection algorithm and the optimal beamformer $\btheta^\star$ satisfy
\begin{equation}
	\label{bound:greedy}
	\cos^2(\pi/K) f(\btheta^\star) \le f(\btheta^{\text{CPP}}) \le f(\btheta^\star).
\end{equation}
\end{proposition}
\begin{IEEEproof}
Clearly, we can upper bound the optimal SNR boost as $f(\bm\theta^\star)\le (1/\beta^2_0)\big(\sum^N_{n=0}\beta_n\big)^2$ by assuming that all the channels can be aligned exactly. We also have
\allowdisplaybreaks
\begin{subequations}
\label{CPP:lower_bound}
\begin{align}
	f(\btheta^\text{CPP})
	&=\frac{1}{\beta^2_0}\cdot\left|\beta_0e^{j\alpha_0}+\sum^N_{n=1}\beta_ne^{j\left(\theta^{\text{CPP}}_n+\alpha_n\right)}\right|^2
		\label{CPP:lower_bound:a}\\
	&=\frac{1}{\beta^2_0}\cdot\left|\beta_0+\sum^N_{n=1}\beta_ne^{j\left(\theta^{\text{CPP}}_n-\theta^\infty_n\right)}\right|^2
		\label{CPP:lower_bound:b}\\
	&\ge\frac{1}{\beta^2_0}\cdot\left|\beta_0+\sum^N_{n=1}\beta_n\cos\left(\theta^{\text{CPP}}_n-\theta^\infty_n\right)\right|^2
		\label{CPP:lower_bound:c}\\
	&\ge\frac{1}{\beta^2_0}\cdot\Bigg|\beta_0+\sum^N_{n=1}\beta_n\cos\frac{\omega}{2}\Bigg|^2
		\label{CPP:lower_bound:d}\\
	&\ge\cos^2(\omega/2)\frac{1}{\beta^2_0}\cdot\Bigg|\sum^N_{n=0}\beta_n\Bigg|^2
		\label{CPP:lower_bound:e}\\
	&\ge \cos^2(\omega/2)f(\btheta^\star),
		\label{CPP:lower_bound:f}
\end{align}
\end{subequations}
where (\ref{CPP:lower_bound:d}) follows since $|\theta^{\text{CPP}}_n-\theta^\infty_n|\le\omega/2$ by the closest point projection. The proof is then completed.
\end{IEEEproof}

\begin{figure}
\begin{minipage}[b]{1.0\linewidth}
	  \centering
	  \centerline{\includegraphics[width=4.4cm]{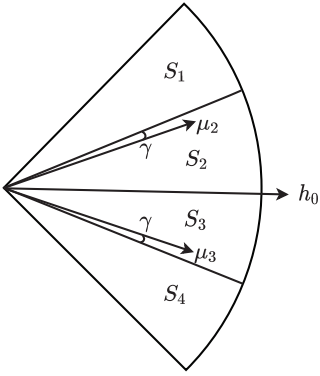}}
\end{minipage}
\caption{The worst-case scenario of the closest point projection algorithm. If $\mu_1=0$ and $|\mu_2|=|\mu_3|>0$, then $f(\btheta^\text{CPP})=(4/\beta^2_0)|\mu_2|^2\cos^2(\pi/K)$ and $f(\btheta^\star)=(4/\beta^2_0)|\mu_2|^2$ as $\gamma=\beta_0=0$, so $f(\btheta^\text{CPP})=\cos^2(\pi/K)f(\btheta^\star)$.}
\label{fig:bad_example}
\end{figure}

Furthermore, we remark that the lower bound in (\ref{bound:greedy}) can be tight as shown in Fig.~\ref{fig:bad_example}.

While the previous works \cite{rui_SDP_BF_19,Johansson_est_19,rui_double_BF_21,Nallanathan_SDP_BF_20,Dobre_SDP_BF_21,YF_maxmin_BF_21,Skoglund_decentralized_BF_21} have used SDR for continuous beamforming, we show that SDR works for discrete beamforming as well. Toward this end, we first rewrite the problem as a complex $K$-ary quadratic problem in (\ref{prob:quadratic:complex}), and then the standard SDR method applies. The next proposition is a direct result from the existing studies on SDR \cite{so_math07,luo_SDR}.

\begin{proposition}
	The $K$-ary phase beamforming problem in (\ref{prob:snr}) can be rewritten as a complex $K$-ary quadratic problem in (\ref{prob:quadratic:complex}); the new problem can be approximately solved by the standard SDR method \cite{so_math07,luo_SDR}. The resulting IRS beamformer $\btheta^{\text{SDR}}$ 
	and the optimal IRS beamformer $\btheta^\star$ satisfy
\begin{equation}
	\frac{(K\sin(\pi/K))^2}{4\pi} f(\btheta^\star) \le f(\btheta^{\text{SDR}}) \le f(\btheta^\star).
\end{equation}
\end{proposition}

\begin{figure}
\begin{minipage}[b]{1.0\linewidth}
	  \centering
	  \centerline{\includegraphics[width=9.7cm]{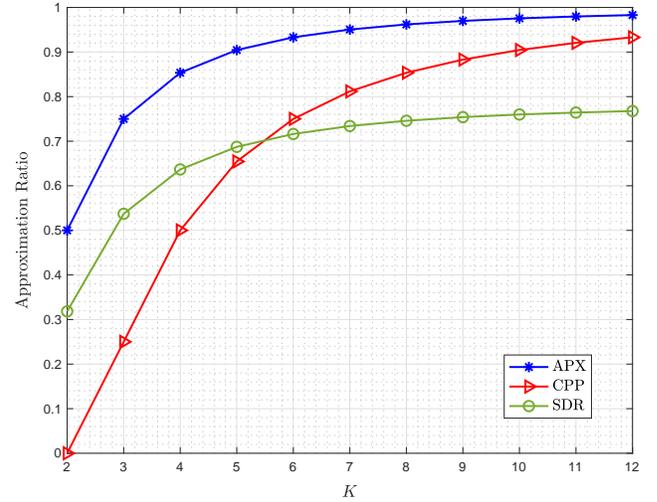}}
\end{minipage}
\caption{Algorithm \ref{alg:K-ary} (APX) vs. closest point projection (CPP) algorithm vs. semidefine relaxation (SDR) algorithm in terms of the approximation ratio.}
\label{fig:approx_ratio_K}
\end{figure}

We conclude this section by comparing the approximation ratios of the various algorithms in Fig.~\ref{fig:approx_ratio_K}. It shows that the guaranteed performance of the proposed approximation algorithm is much closer to the global optimum.

\section{Simulation Results}
\label{sec:simulations}

\begin{figure}
\begin{minipage}[b]{1.0\linewidth}
	  \centering
	  \centerline{\includegraphics[width=9.7cm]{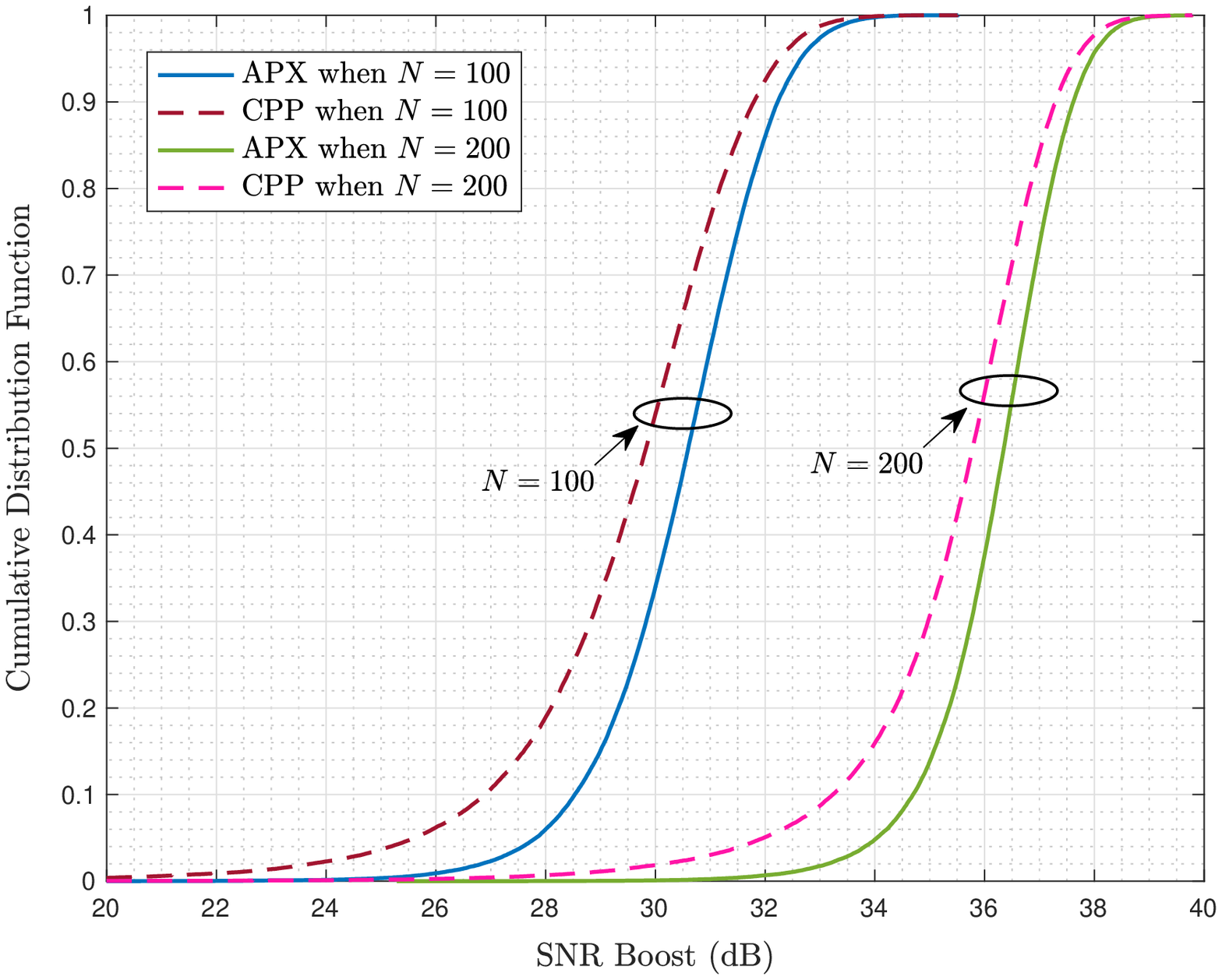}}
\end{minipage}
\caption{The cumulative distribution of the SNR boost when $K=2$.}
\label{fig:CDF_K2}
\end{figure}

\begin{figure}
\begin{minipage}[b]{1.0\linewidth}
	  \centering
	  \centerline{\includegraphics[width=9.7cm]{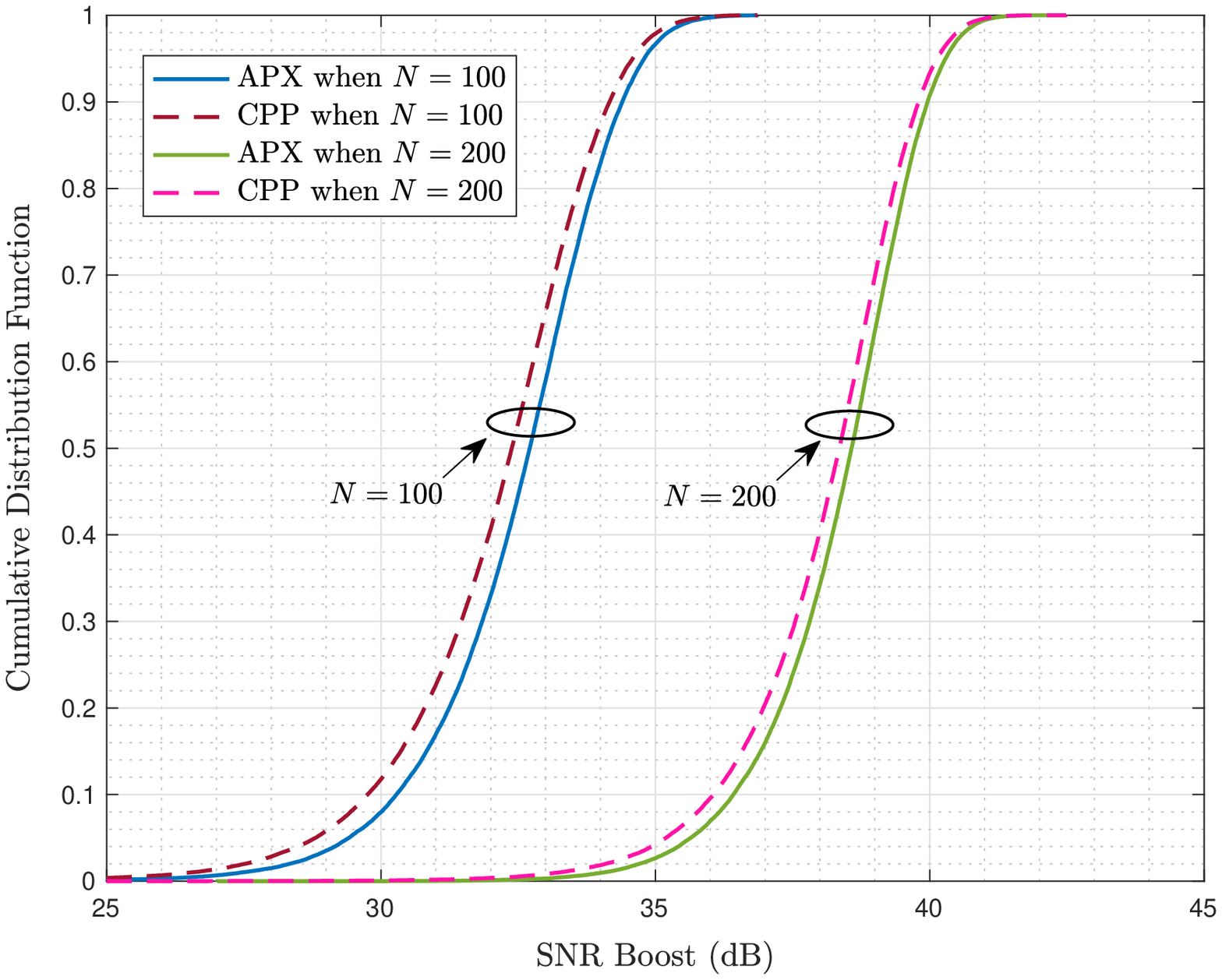}}
\end{minipage}
\caption{The cumulative distribution of the SNR boost when $K=4$.}
\label{fig:CDF_K4}
\end{figure}

\begin{figure}
\begin{minipage}[b]{1.0\linewidth}
	  \centering
	  \centerline{\includegraphics[width=9.7cm]{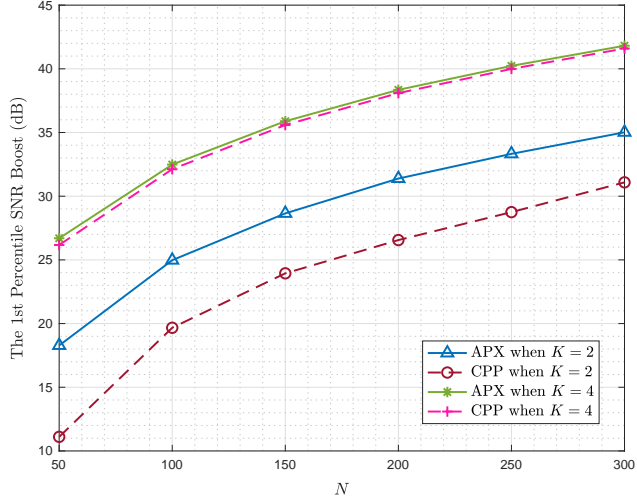}}
\end{minipage}
\caption{The number of reflective elements $N$ vs. the 1st percentile SNR boost.}
\label{fig:N_percentile_boost}
\end{figure}

\begin{figure}
\begin{minipage}[b]{1.0\linewidth}
	  \centering
	  \centerline{\includegraphics[width=9.7cm]{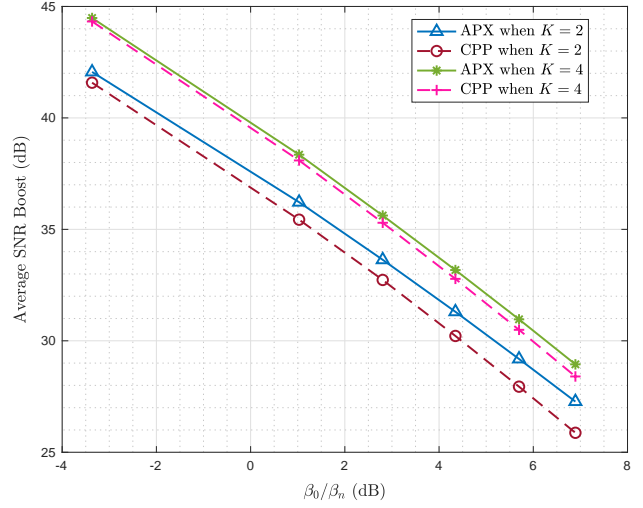}}
\end{minipage}
\caption{The background-to-reflection ratio $\beta_0/\beta_n$ vs. the average SNR boost.}
\label{fig:beta_boost}
\end{figure}

\begin{figure}
\begin{minipage}[b]{1.0\linewidth}
	  \centering
	  \centerline{\includegraphics[width=9.7cm]{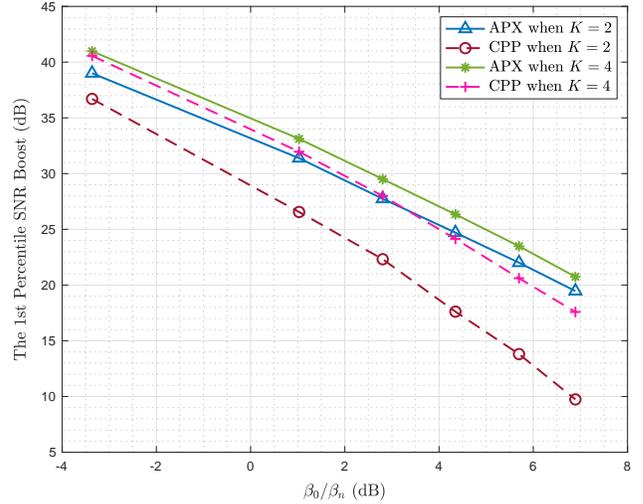}}
\end{minipage}
\caption{The channel ratio $\beta_0/\beta_n$ vs. the 1st percentile SNR boost.}
\label{fig:beta_percentile_boost}
\end{figure}

In this section we validate the performance of the proposed algorithms in simulations. The channel model follows the previous works \cite{rui_SDP_BF_19,Yuen_DBF_20,wy_AI_21}. The background channel $h_0$ is given by
\begin{equation}
	h_0 = 10^{-\mathsf{PL}_0/20}\cdot\zeta_0,
\end{equation}
where the transmitter-to-receiver pathloss $\mathsf{PL}_0$ (in dB) is computed as $\mathsf{PL}_0=32.6+36.7\log_{10}(d_0)$, with $d_0$ in meters denoting the distance between the transmitter and the receiver, while the Rayleigh fading component $\zeta_0$ is drawn from the Gaussian distribution $\mathcal{CN}(0,1)$. The reflected channel $h_n$ 
is given by
\begin{equation}
	h_n = 10^{-(\mathsf{PL}_1+\mathsf{PL}_2)/20}\cdot\zeta_n, n=1,\ldots,N,
\end{equation}
where $\mathsf{PL}_1$ and $\mathsf{PL}_2$ are both based on the pathloss model $\mathsf{PL}_0=30+22\log_{10}(d)$, with $d$ in meters respectively denoting the transmitter-to-IRS distance and the IRS-to-receiver distance, while the Rayleigh fading component $\zeta_n$ is drawn from the Gaussian distribution $\mathcal{CN}(0,1)$ independently across $n=1,\ldots,N$. We use the following parameters unless otherwise stated. The transmit power level $P$ equals $30$ dBm. The background noise power level $\sigma^2$ equals $-90$ dBm. The locations of the transmitter, IRS, and receiver are denoted by the 3-dimensional coordinate vectors $(50,-200,20)$, $(-2,-1,0)$, and $(0,0,0)$ in meters, respectively. The number of reflective elements $N$ equals $200$. The channels are estimated via an {\footnotesize{ON-OFF}} strategy in \cite{Carvalho_est_20}.

Fig.~\ref{fig:CDF_K2} shows the cumulative distribution of the SNR boost for the binary phase beamforming case with $K=2$. When $N=100$, it can be seen that the proposed approximation algorithm ``APX'' outperforms the closest point projection algorithm ``CPP'', especially in the low SNR boost regime. For instance, the 5th percentile SNR boost by APX is more than 2 dB higher than that by CPP. This performance gap remains when the number of reflective elements raises to 200. Moreover, we consider the quadrature beamforming case with $K=4$ in Fig.~\ref{fig:CDF_K4}. Observe that APX and CPP now have similar performance. Thus, APX is more suited for the binary phase beamforming for IRS.





We further look into the low SNR boost regime by comparing the 1st percentile SNR boosts (i.e., the 1\% lowest SNR boost across the random channel realizations) in Fig.~\ref{fig:N_percentile_boost}. Different values of $N$ are considered here. The figure shows that APX outperforms CPP significantly in the low SNR boost regime when $K=2$. For instance, when $N=200$, APX improves upon CPP by about 5 dB. Observe that the gap between APX and CPP decreases with $N$. When $K$ is increased to $4$, the advantage of APX over CPP becomes marginal.

Fig.~\ref{fig:beta_boost} shows how the reflected channel strength impacts the SNR boost. We put the IRS sequentially at the following 6 positions: $(-1,-1,0)$, $(-2,-1,0)$, $(-2.5,-1,0)$, $(-3,-1,0)$, $(-3.5,-1,0)$, $(-4,-1,0)$, so the reflected channel magnitude $\beta_n$ varies among these IRS locations, while the background channel magnitude $\beta_0$ is fixed. We evaluate the channel ratio $\beta_0/\beta_n$ for each IRS location. As shown in Fig.~\ref{fig:beta_boost}, the gap between APX and CPP in terms of the average SNR boost increases with $\beta_0/\beta_n$, in both the binary phase beamforming case and the quadrature beamforming case. Furthermore, Fig.~\ref{fig:beta_percentile_boost} shows the low SNR boost performance with respect to the different values of $\beta_0/\beta_n$. The figure also suggests that the gap between APX and CPP becomes larger when $\beta_0/\beta_n$ is increased. Summarizing the above observations, we conclude that APX is more suited for the scenario where the reflected channels are relatively weak in comparison to the background channel.

\begin{figure}
\begin{minipage}[b]{1.0\linewidth}
	  \centering
	  \centerline{\includegraphics[width=9.7cm]{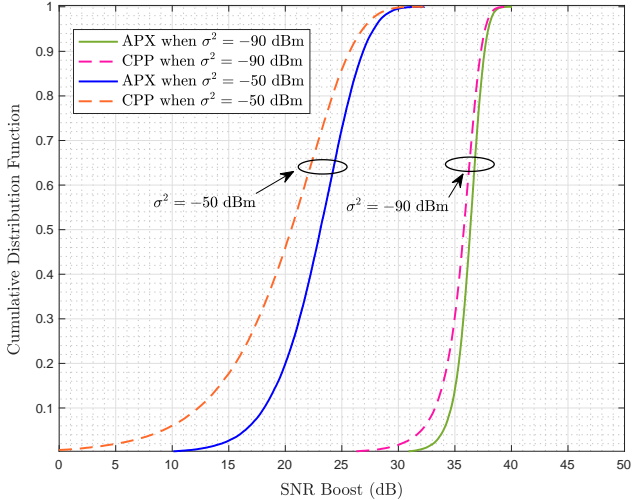}}
\end{minipage}
\caption{The cumulative distributions of the SNR boost under the different noise powers when $K=2$.}
\label{fig:CDF_noise}
\end{figure}

The noise power $\sigma^2$ is fixed at $-90$ dBm in all the above simulations. We now raise $\sigma^2$ to $-50$ dBm and plot the resulting cumulative distributions of SNR boosts in Fig.~\ref{fig:CDF_noise}. Due to more severe noise, the channel estimation becomes less accurate, and thus can spoil the IRS beamforming algorithms. Observe from Fig.~\ref{fig:CDF_noise} that APX is less affected by the increased noise as compared to CPP. Actually,
APX is far superior to CPP when $\sigma^2=-50$ dBm, whereas the two algorithms are close when $\sigma^2=-90$ dBm. The tolerance to channel estimation error enables APX to perform much better than CPP in the presence of strong interference and background noise.

\section{Conclusion}
\label{sec:conclusion}
Despite the practical discrete phase value constraints for the IRS beamforming, this work shows that the optimal SNR boost can be reached by the proposed algorithm with a linear time complexity in the number of reflective elements, if the phase value is binary. For the general $K$-ary phase beamforming, we propose a linear time algorithm with a provable approximation accuracy, whereas the existing closest point projection algorithm can result in arbitrarily bad performance. Simulation results demonstrate the superior performance of  the proposed approximation algorithm over the existing closest point projection algorithm in enhancing the received SNR, especially when channel estimation is error-prone because of interference and noise.



\bibliographystyle{IEEEtran}     
\bibliography{IEEEabrv,strings}     

\end{document}